\documentclass{psa}

\usepackage{amsmath,amsthm,dsfont,enumitem,mathtools}
\usepackage{mdframed}
\usepackage{orcidlink}

\usepackage[normalem]{ulem}
\usepackage{csquotes}
\setcitestyle{authoryear,open={(},close={)}}

\begin{document}

\doival{10.1017/xxxxx}
\jnlPage{x}{x}
\jnlDoiYr{2026}
\lefttitle{Cafiero, Molinari, \& Hance}

\papertitle{RESEARCH ARTICLE}

\title{The Burden of Fundamentality: Metaphysical Ambiguities and the Issue of Superdeterminism}

\author{Gabriele Cafiero$^{1}$, Luca Molinari$^1$ and Jonte R. Hance$^1$\,\orcidlink{0000-0001-8587-7618}}

\affil{$^{1}$Quantum Group, School of Computing, Newcastle University, 1 Science Square, Newcastle upon Tyne, NE4 5TG, UK\\
      $^\ast$Corresponding author: Jonte R. Hance;
      \email{jonte.hance@newcastle.ac.uk}}

\received{21 January 2026}
\revised{23 June 2026}
\accepted{01 August 2026}

\begin{abstract}
In this paper we approach the problem of superdeterminism from a novel point of view, highlighting its character as a more metaphysical than scientific proposition. First, we introduce a distinction between two types of superdeterministic theories, \textit{na\"ive} (NSD) and \textit{metaphysical} (MSD), and argue how NSD presents significant epistemic flaws. We show how NSD justifies itself through claims to fundamentality, thus connoting itself as a metaphysical theory rather than a scientific one. We finally illustrate that the most developed MSD model so far, Invariant Set Theory, implicitly proposes a confused form of priority monism. 
Our paper thus reinforces the thesis that theories should demonstrate rather than assume fundamentality and that it is methodologically flawed for a theory to assume its own fundamentality for the sole purpose of defending against criticisms.
\end{abstract}


\maketitle

\section{Introduction}\label{sec:Intro}

Superdeterminism has taken a prominent role in recent debates on the assumptions and implications of Bell’s theorem. Although latest contributions have developed this notion from its reductio-ad-absurdum roots~\citep{HossenfelderPalmer2020,Andreoletti2022,HanceHossenfelder2022,Hance2024}, critiques of superdeterminism have typically recycled older arguments~\citep{Maudlin2019,Chen2021}, with a few notable exceptions~\citep{Baas2023}. The main criticisms against rejecting statistical independence still tend to focus either on concerns regarding risks to scientific practice~\citep{Maudlin2019} or the supposed implausibility, fine-tuned-ness, and conspiratorial nature of models that accommodate superdeterminism~\citep{Chen2021,Daley2022Adjudicating,Hance2024DaleyComment,Daley2024HanceReply}. Throughout this body of literature, there is an implicit effort to discredit superdeterminism as an implausible or even dangerous scientific theory, leading to a kind of argumentative stalemate: given that superdeterminism is logically viable, a determined superdeterminist can respond to nearly any criticism by addressing specific implausibilities and risks~\citep{Andreoletti2022}.

This paper instead disambiguates superdeterministic models into three categories. We show that the first of these, which we term \textit{na\"ive superdeterminism}, relies on illegitimate claims to fundamentality---thereby presenting itself more as a metaphysical theory than of a scientific one. We then look at the second category, \textit{metaphysical superdeterminism}, which typically built from models from the first category to legitimise claims of fundamentality. We show that the most well-developed example of a \textit{metaphysical superdeterministic} model, Invariant Set Theory (IST), displays a degree of metaphysical conceptual confusion that renders it unacceptable.

To this end, the paper is organized as follows: in Sec.~\ref{sec:WhatIs?} and Sec.~\ref{sec:Whatare}, we will briefly recall what is meant by superdeterminism, and the theoretical problem it has been invoked to address, before differentiating it into three categories. In Sec.~\ref{sec:Metaphys} we will show that the first and second category of superdeterminism makes illegitimate use of claims to fundamentality, thus exhibiting more the character of a metaphysical theory rather than a scientific one. In Sec.~\ref{sec:MetaSoln}, we will present the features of this metaphysical theory through an analysis of IST. In Sec.~\ref{sec:Ambig}, we will demonstrate that IST presents itself as a highly confused metaphysical framework. In Sec.~\ref{sec:Disc}, we will draw some general conclusions.

\section{What is Superdeterminism?}\label{sec:WhatIs?}

\begin{figure}
    \centering

    \begin{tikzpicture}[x=0.75pt,y=0.75pt,yscale=-1,xscale=1]

\draw    (222,37) -- (222,220.7) ;
\draw [shift={(222,34)}, rotate = 90] [fill={rgb, 255:red, 0; green, 0; blue, 0 }  ][line width=0.08]  [draw opacity=0] (8.93,-4.29) -- (0,0) -- (8.93,4.29) -- cycle    ;
\draw    (450.43,210.7) -- (213.43,210.7) ;
\draw [shift={(453.43,210.7)}, rotate = 180] [fill={rgb, 255:red, 0; green, 0; blue, 0 }  ][line width=0.08]  [draw opacity=0] (8.93,-4.29) -- (0,0) -- (8.93,4.29) -- cycle    ;
\draw   (317.2,175.35) .. controls (317.2,165.6) and (325.1,157.7) .. (334.85,157.7) .. controls (344.6,157.7) and (352.5,165.6) .. (352.5,175.35) .. controls (352.5,185.1) and (344.6,193) .. (334.85,193) .. controls (325.1,193) and (317.2,185.1) .. (317.2,175.35) -- cycle ;
\draw [color={rgb, 255:red, 208; green, 2; blue, 27 }  ,draw opacity=1 ]   (291.19,130.99) -- (322.43,162.23) ;
\draw [shift={(289.07,128.87)}, rotate = 45] [fill={rgb, 255:red, 208; green, 2; blue, 27 }  ,fill opacity=1 ][line width=0.08]  [draw opacity=0] (8.93,-4.29) -- (0,0) -- (8.93,4.29) -- cycle    ;
\draw [color={rgb, 255:red, 208; green, 2; blue, 27 }  ,draw opacity=1 ]   (348.2,163) -- (378.95,132.25) ;
\draw [shift={(381.07,130.13)}, rotate = 135] [fill={rgb, 255:red, 208; green, 2; blue, 27 }  ,fill opacity=1 ][line width=0.08]  [draw opacity=0] (8.93,-4.29) -- (0,0) -- (8.93,4.29) -- cycle    ;
\draw   (263.2,118.83) .. controls (263.2,110.46) and (269.99,103.67) .. (278.37,103.67) .. controls (286.74,103.67) and (293.53,110.46) .. (293.53,118.83) .. controls (293.53,127.21) and (286.74,134) .. (278.37,134) .. controls (269.99,134) and (263.2,127.21) .. (263.2,118.83) -- cycle ;
\draw    (278.37,71.67) -- (278.37,103.67) ;
\draw [shift={(278.37,68.67)}, rotate = 90] [fill={rgb, 255:red, 0; green, 0; blue, 0 }  ][line width=0.08]  [draw opacity=0] (8.93,-4.29) -- (0,0) -- (8.93,4.29) -- cycle    ;
\draw   (375.2,117.83) .. controls (375.2,109.46) and (381.99,102.67) .. (390.37,102.67) .. controls (398.74,102.67) and (405.53,109.46) .. (405.53,117.83) .. controls (405.53,126.21) and (398.74,133) .. (390.37,133) .. controls (381.99,133) and (375.2,126.21) .. (375.2,117.83) -- cycle ;
\draw    (390.37,70.67) -- (390.37,102.67) ;
\draw [shift={(390.37,67.67)}, rotate = 90] [fill={rgb, 255:red, 0; green, 0; blue, 0 }  ][line width=0.08]  [draw opacity=0] (8.93,-4.29) -- (0,0) -- (8.93,4.29) -- cycle    ;
\draw   (263.2,53.5) .. controls (263.2,45.12) and (269.99,38.33) .. (278.37,38.33) .. controls (286.74,38.33) and (293.53,45.12) .. (293.53,53.5) .. controls (293.53,61.88) and (286.74,68.67) .. (278.37,68.67) .. controls (269.99,68.67) and (263.2,61.88) .. (263.2,53.5) -- cycle ;
\draw   (375.2,52.5) .. controls (375.2,44.12) and (381.99,37.33) .. (390.37,37.33) .. controls (398.74,37.33) and (405.53,44.12) .. (405.53,52.5) .. controls (405.53,60.88) and (398.74,67.67) .. (390.37,67.67) .. controls (381.99,67.67) and (375.2,60.88) .. (375.2,52.5) -- cycle ;
\draw    (222,247.67) -- (222,431.37) ;
\draw [shift={(222,244.67)}, rotate = 90] [fill={rgb, 255:red, 0; green, 0; blue, 0 }  ][line width=0.08]  [draw opacity=0] (8.93,-4.29) -- (0,0) -- (8.93,4.29) -- cycle    ;
\draw    (450.43,421.37) -- (213.43,421.37) ;
\draw [shift={(453.43,421.37)}, rotate = 180] [fill={rgb, 255:red, 0; green, 0; blue, 0 }  ][line width=0.08]  [draw opacity=0] (8.93,-4.29) -- (0,0) -- (8.93,4.29) -- cycle    ;
\draw   (317.2,386.02) .. controls (317.2,376.27) and (325.1,368.37) .. (334.85,368.37) .. controls (344.6,368.37) and (352.5,376.27) .. (352.5,386.02) .. controls (352.5,395.76) and (344.6,403.67) .. (334.85,403.67) .. controls (325.1,403.67) and (317.2,395.76) .. (317.2,386.02) -- cycle ;
\draw [color={rgb, 255:red, 0; green, 0; blue, 0 }  ,draw opacity=1 ]   (294.64,266.96) -- (334.85,368.37) ;
\draw [shift={(293.53,264.17)}, rotate = 68.37] [fill={rgb, 255:red, 0; green, 0; blue, 0 }  ,fill opacity=1 ][line width=0.08]  [draw opacity=0] (8.93,-4.29) -- (0,0) -- (8.93,4.29) -- cycle    ;
\draw [color={rgb, 255:red, 0; green, 0; blue, 0 }  ,draw opacity=1 ]   (334.85,368.37) -- (374.13,265.97) ;
\draw [shift={(375.2,263.17)}, rotate = 110.98] [fill={rgb, 255:red, 0; green, 0; blue, 0 }  ,fill opacity=1 ][line width=0.08]  [draw opacity=0] (8.93,-4.29) -- (0,0) -- (8.93,4.29) -- cycle    ;
\draw   (263.2,329.5) .. controls (263.2,321.12) and (269.99,314.33) .. (278.37,314.33) .. controls (286.74,314.33) and (293.53,321.12) .. (293.53,329.5) .. controls (293.53,337.88) and (286.74,344.67) .. (278.37,344.67) .. controls (269.99,344.67) and (263.2,337.88) .. (263.2,329.5) -- cycle ;
\draw   (375.2,328.5) .. controls (375.2,320.12) and (381.99,313.33) .. (390.37,313.33) .. controls (398.74,313.33) and (405.53,320.12) .. (405.53,328.5) .. controls (405.53,336.88) and (398.74,343.67) .. (390.37,343.67) .. controls (381.99,343.67) and (375.2,336.88) .. (375.2,328.5) -- cycle ;
\draw   (263.2,264.17) .. controls (263.2,255.79) and (269.99,249) .. (278.37,249) .. controls (286.74,249) and (293.53,255.79) .. (293.53,264.17) .. controls (293.53,272.54) and (286.74,279.33) .. (278.37,279.33) .. controls (269.99,279.33) and (263.2,272.54) .. (263.2,264.17) -- cycle ;
\draw   (375.2,263.17) .. controls (375.2,254.79) and (381.99,248) .. (390.37,248) .. controls (398.74,248) and (405.53,254.79) .. (405.53,263.17) .. controls (405.53,271.54) and (398.74,278.33) .. (390.37,278.33) .. controls (381.99,278.33) and (375.2,271.54) .. (375.2,263.17) -- cycle ;
\draw [color={rgb, 255:red, 208; green, 2; blue, 27 }  ,draw opacity=1 ][fill={rgb, 255:red, 208; green, 2; blue, 27 }  ,fill opacity=1 ]   (289.83,339.65) -- (321.08,370.9) ;
\draw [shift={(323.2,373.02)}, rotate = 225] [fill={rgb, 255:red, 208; green, 2; blue, 27 }  ,fill opacity=1 ][line width=0.08]  [draw opacity=0] (8.93,-4.29) -- (0,0) -- (8.93,4.29) -- cycle    ;
\draw [color={rgb, 255:red, 208; green, 2; blue, 27 }  ,draw opacity=1 ]   (350.32,371.88) -- (381.07,341.13) ;
\draw [shift={(348.2,374)}, rotate = 315] [fill={rgb, 255:red, 208; green, 2; blue, 27 }  ,fill opacity=1 ][line width=0.08]  [draw opacity=0] (8.93,-4.29) -- (0,0) -- (8.93,4.29) -- cycle    ;
\draw [color={rgb, 255:red, 0; green, 0; blue, 0 }  ,draw opacity=1 ][fill={rgb, 255:red, 0; green, 0; blue, 0 }  ,fill opacity=1 ]   (294.64,56.29) -- (334.85,157.7) ;
\draw [shift={(293.53,53.5)}, rotate = 68.37] [fill={rgb, 255:red, 0; green, 0; blue, 0 }  ,fill opacity=1 ][line width=0.08]  [draw opacity=0] (8.93,-4.29) -- (0,0) -- (8.93,4.29) -- cycle    ;
\draw [color={rgb, 255:red, 0; green, 0; blue, 0 }  ,draw opacity=1 ]   (334.85,157.7) -- (374.13,55.3) ;
\draw [shift={(375.2,52.5)}, rotate = 110.98] [fill={rgb, 255:red, 0; green, 0; blue, 0 }  ,fill opacity=1 ][line width=0.08]  [draw opacity=0] (8.93,-4.29) -- (0,0) -- (8.93,4.29) -- cycle    ;
\draw    (278.37,282.33) -- (278.37,314.33) ;
\draw [shift={(278.37,279.33)}, rotate = 90] [fill={rgb, 255:red, 0; green, 0; blue, 0 }  ][line width=0.08]  [draw opacity=0] (8.93,-4.29) -- (0,0) -- (8.93,4.29) -- cycle    ;
\draw    (390.37,281.33) -- (390.37,313.33) ;
\draw [shift={(390.37,278.33)}, rotate = 90] [fill={rgb, 255:red, 0; green, 0; blue, 0 }  ][line width=0.08]  [draw opacity=0] (8.93,-4.29) -- (0,0) -- (8.93,4.29) -- cycle    ;

\draw (177.2,25) node [anchor=north west][inner sep=0.75pt]   [align=left] {$\displaystyle \text{time}$};
\draw (436.2,218) node [anchor=north west][inner sep=0.75pt]   [align=left] {$\displaystyle \text{space}$};
\draw (329.8,168.67) node [anchor=north west][inner sep=0.75pt]   [align=left] {$\displaystyle \lambda $};
\draw (273.47,113.67) node [anchor=north west][inner sep=0.75pt]   [align=left] {$\displaystyle a$};
\draw (387.47,110.67) node [anchor=north west][inner sep=0.75pt]   [align=left] {$\displaystyle b$};
\draw (271.47,45.67) node [anchor=north west][inner sep=0.75pt]   [align=left] {$\displaystyle A$};
\draw (385.47,45.67) node [anchor=north west][inner sep=0.75pt]   [align=left] {$\displaystyle B$};
\draw (177.2,235.67) node [anchor=north west][inner sep=0.75pt]   [align=left] {$\displaystyle \text{time}$};
\draw (436.2,428.67) node [anchor=north west][inner sep=0.75pt]   [align=left] {$\displaystyle \text{space}$};
\draw (329.8,379.33) node [anchor=north west][inner sep=0.75pt]   [align=left] {$\displaystyle \lambda $};
\draw (273.47,325.33) node [anchor=north west][inner sep=0.75pt]   [align=left] {$\displaystyle a$};
\draw (387.47,323.33) node [anchor=north west][inner sep=0.75pt]   [align=left] {$\displaystyle b$};
\draw (270.47,256.33) node [anchor=north west][inner sep=0.75pt]   [align=left] {$\displaystyle A$};
\draw (384.47,256.33) node [anchor=north west][inner sep=0.75pt]   [align=left] {$\displaystyle B$};
\draw (182,186) node [anchor=north west][inner sep=0.75pt]   [align=left] {a.};
\draw (186,399) node [anchor=north west][inner sep=0.75pt]   [align=left] {b.};

\end{tikzpicture}

    \caption{Two different SI-violating models, for a bipartite scenario where the two parts are spacelike separated (e.g., the CHSH scenario~\citep{ClauserShimonyHolt1969}). Arrows in black indicate causal dependencies common between the model and a Statistical Independence-preserving Local Hidden Variable Model; red arrows indicate the causal dependencies by which the model violates Statistical Independence. \textbf{a.} Superdeterministic models: the hidden variables ($\lambda$ determines the measurement settings (\textit{a} and \textit{b}). \textbf{b.} Future Input-Dependent models: the measurement settings $a$ and $b$ retrocausally determine the hidden variables ($\lambda$). This means that mathematically $\lambda$ is correlated with $a$ and $b$, while physically it means that in some sense---which differs depending on the model---the settings $a$ and $b$ cause $\lambda$.{In every model, however, information transferring backward in time is not \textit{accessible} or \textit{usable}; in other words, retrocausal signalling is typically not allowed.}}
    \label{fig:FID}
\end{figure}
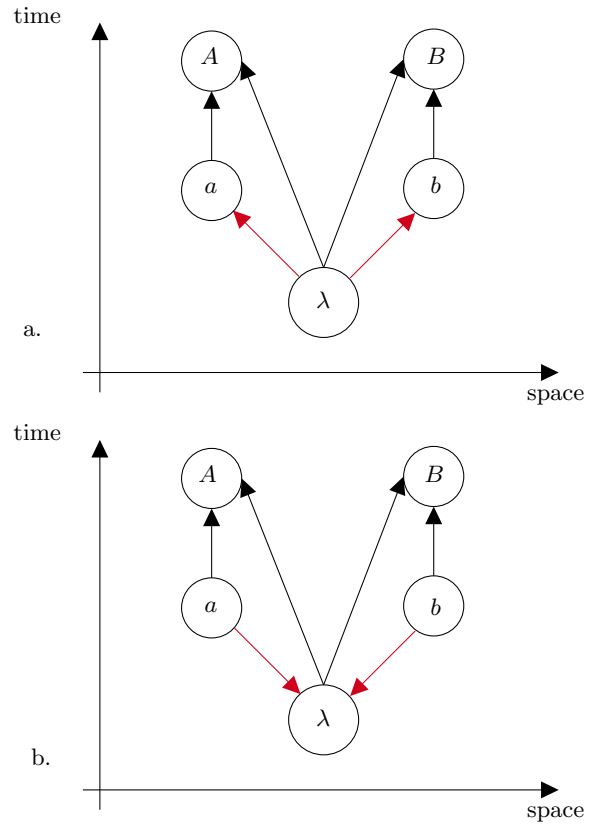

{Superdeterminism refers to a} class of local hidden-variable deterministic interpretations (or extensions~\citep[p.5]{adlam2023taxonomy}) of quantum mechanics whose primary feature is the violation of the statistical independence assumption (SI){~{\citep{HossenfelderPalmer2020,Andreoletti2022}}. In practical terms, this means that the probability distribution of hidden variables $\rho(\lambda)$ is not independent of the choice of measurement. Specifically, if $a$ denotes the measurement choice and $\lambda$ the hidden variables:}
\begin{equation}\label{eq:SI}
    \rho(\lambda \mid a) \neq \rho(\lambda)
\end{equation}

{Violating SI is, however, a necessary but not sufficient condition for superdeterminism. Other models---such as Future Input Dependent (FID) models~{\citep{Wharton2019Reformulations,Waegell2025}}---also violate SI without qualifying as superdeterministic. The key distinction lies in the underlying physical mechanism. In FID models, $\lambda$---though located in the past light cone of the measurement settings---is in some way constrained by those settings. In superdeterminism, by contrast, the measurement settings themselves are determined or constrained by $\lambda$.}

Existing attempts to capture this distinction formally are not without controversy. \cite{adlam2023taxonomy} exclude retrocausal approaches from being considered superdeterminism simply on the grounds that they have traditionally never been considered superdeterministic, while \cite{Waegell2025} exclude them on the basis of a merely temporal violation of SI---a criterion that would exclude theories such as Invariant Set Theory.

{A more satisfactory definition can be recovered by returning to Bell's theorem itself, from which superdeterminism originates. Proponents of superdeterminism propose violating SI rather than accepting nonlocality, thereby preserving the intuition that there are no \textit{spooky actions at a distance}. This is possible precisely because SI and locality are logically independent premises in the derivation of Bell inequalities. Bell's theorem establishes that, under certain classical and sufficiently reasonable assumptions, one obtains inequalities whose violation by quantum mechanics forces the abandonment of at least one such assumption~{\citep{Bell1964,EinsteinPodolskyRosen1935}}.} 

{Coming back to our problem of distinguishing superdeterminism from retrocausal and other SI-violating approaches, we will employ the causal-model approach applied to Bell's theorem by \cite{Wharton2019Reformulations}, modelled after the framework introduced by \cite{pearl2009causality}. Within this framework, in the context of a CHSH experiment, the measurement settings $a$ and $b$ can be treated either as free input parameters or as stochastic variables. When treated as free inputs, one recovers the \textit{measurement independence} condition, $P_c(\lambda \mid a, b) = P_c(\lambda)$, which follows from the assumption of No Future Input Dependence (NFID)}\footnote{Meaning that the evaluation of events up to a time $t'$, does not rely upon events for $t>t'$ \citep[p.13]{Wharton2019Reformulations}.}. Superdeterministic models violate this condition by treating $a$ and $b$ as stochastic variables rather than free inputs. Then, through a Bayesian inversion of the measurement independence condition, the \textit{no-conspiracies} assumption follows,
\begin{equation}
    P_c(a, b \mid \lambda) = P_c(a, b),
\end{equation}
which superdeterministic models violate. Retrocausal models, by contrast, violate NFID itself, and are therefore not superdeterministic in this sense. We can thus obtain the following definition:

\begin{definition}
  Superdeterminism is a class of local hidden-variable deterministic interpretations (or extensions) of quantum mechanics whose primary feature is the treatment of measurement settings $a$ and $b$ as statistical parameters, and the concomitant violation of the no-conspiracies assumption.
\end{definition}

\section{What are Superdeterminisms?}
\label{sec:Whatare}

If it is true that a necessary condition for defining a superdeterministic theory is the violation of SI, it is equally true that not all superdeterministic theories restrict themselves to this feature alone. Attempts to classify the various superdeterministic proposals are certainly not lacking~\citep{Waegell2025,SenValentini2020,SenValentini2020b}. For our purposes, we shall consider three categories of superdeterminism:
\begin{enumerate}

    \item\label{Category:CSD} The first category is what we may call \textit{na\"ive superdeterminism} (NSD). Stemming from Brans' model~\citep{Brans1988Model}, NSD is the superdeterministic position represented, for example, by \cite{Andreoletti2022} or criticised by \cite{SenValentini2020,SenValentini2020b}. As we shall show in Section~\ref{sec:Metaphys}, in order to defend themselves from extant {epistemological} criticisms, models in this category present themselves as \textit{fundamental}.
    
    \item\label{Category:MSD} The second category is what we may call \textit{metaphysical} or \textit{less-na\"ive superdeterminism} (MSD). MSD models overcome the epistemic difficulties we discuss in Section~\ref{sec:Metaphys}, and consist of models constructed to fit the criteria for fundamentality that NSD models simply assume they {possess, thus fulfilling the promise introduced by NSD models.} Examples of this category include IST (discussed in Section~\ref{sec:MetaSoln}) and 't Hooft's Cellular Automaton model~(\citeyear{tHooft2016cellular}).
    
    \item\label{Category:PSIV} The third category is represented by physically credible toy-models of SI violation, which we'll call Toy Superdeterminism (TSD). The main difference between models in this category and those in either NSD or MSD is that they are explicitly designed as toy models, to reproduce some aspect of the physics of a certain situation, without any claims as to being the basis of a full-fledged physical theory, and thus without any claims of fundamentality. Their development is tied to an exploratory role (to use the terminology of \cite{Fisher2021}) within the superdeterministic framework. This means that they avoid the difficulties of \eqref{Category:CSD}, without implying different metaphysics as in \eqref{Category:MSD} and therefore we won't consider them further in this paper\footnote{\cite{GarzaHance2025} for example propose that the conservation of symmetries according to Wigner-Araki-Yanase could act as a bias on measurements, restricting some counterfactuals (and thus modifying the measurement space). This presents a physical motivation for the model being supermeasured \citep{HanceHossenfelderPalmer2022}, leaving it debatable whether it causally violates SI.}.
    
\end{enumerate}

The main misunderstanding in previous literature has been to treat superdeterministic theories as having the properties of both NSD and MSD, without further clarifying the distinction between these classes. In fact, the domains in which these two proposals are situated are structurally different. As we shall see, the problem of NSD is primarily \textit{epistemological} (namely, the existence of claims to fundamentality within its theoretical construction). In contrast, the problem of MSD is \textit{metaphysical}, i.e. that the instances seen so far of (this form of) superdeterminism seem to involve a confused and unpalatable metaphysics. It is therefore not possible to develop a single, unified approach to the issue. In a certain sense, our thesis may be divided into two separate claims:
\begin{enumerate}
    \item A NSD theory must necessarily present itself as fundamental in order to shield itself from objections, thereby connoting itself as a theory that is more metaphysical than scientific (as we argue in Section~\ref{sec:Metaphys});
    \item The metaphysical theory constructed on the basis of NSD's requirements (MSD) is confused and unpalatable. We conjecture this in Section~\ref{sec:MetaSoln}, and we show it formally for IST, the most well-developed example of MSD so far.
\end{enumerate}
Of course, not every proposal for SD can be neatly classified into one category or the other. Invariant Set Theory, which we discuss at length in Section~{\ref{sec:MetaSoln}}, for example, could instead be interpreted as a nai\"ve superdeterministic model, if we simply take it as assuming (rather than arguing for) its fundamentality. However, this leads to a different set of problems: the metaphysical superdeterministic interpretation of IST, in actively constructing a fundamental metaphysical structure, overcomes the problems faced by its nai\"ve superdeterministic interpretation, at the expense of gaining the problems associated with metaphysical superdeterminism.

\section{An Epistemological Problem}
\label{sec:Metaphys}

Having explored what is meant by superdeterminism, we will now show how \textit{na\"ive superdeterminism} makes illegitimate use of claims to fundamentality within its theoretical structure. To this end, we will analyse \textit{na\"ive superdeterminism}'s response to two different types of objections: the objection regarding the laws of nature~\citep{Chen2021,Baas2023} and the objection concerning the impossibility of science~\citep{ClauserShimonyHolt1969,Maudlin2019}. However, this analysis could readily apply to its response to all objections raised in the literature, given such responses also typically rely on assuming fundamentality \citep{HossenfelderPalmer2020,Andreoletti2022,Ciepielewski2023}.

\subsection{The Objection Concerning the Laws of Nature}
\label{subsec:LoNObjn}

The objection concerning the laws of nature has received two distinct formulations. The primary one, proposed by \cite{Chen2021}, argues that in a superdeterministic theory, fundamental laws, depending on initial conditions, exhibit a level of complexity not typically attributed to laws considered to be fundamental. This formulation is further strengthened by Baas \& Le Bihan, according to whom, in superdeterminism:
\begin{displayquote}
    \textit{``First, the laws of quantum mechanics are contingent---indeed, in possible worlds with different initial conditions, these laws do not exist. Second, and more problematically, these laws ontologically depend upon the initial conditions. The unfortunate conclusion would then be the possibility of admitting superdeterminism only by assuming a Neo-Humean doctrine of physical laws.''}~\citep[p.12]{Baas2023}
\end{displayquote}

The most comprehensive response to this objection is by \cite{Andreoletti2022}. According to them, the error made by Chen, and Baas \& Le Bihan, is in identifying superdeterminism as merely an interpretation of QM. By contrast, the NSD project aims precisely to construct a more fundamental theory, one capable of explaining the behaviour of the hidden variables $\lambda$ by determining their dynamic evolution. The relevant analogy here would then be statistical mechanics, where the mechanical properties of each particle within a hugely many-particle system are considered to exist, and to be {fully determined by a limited set of simple fundamental laws governing their interactions. At a macroscopic level, the behaviour of each and every particle in the system is not fully knowable in complete detail, and so we replace the variables we would use to describe the behaviour of these particles (e.g., a position and momentum for every particle) with a smaller set of more manageable variables, such as the partition function---or, in superdeterminism's case, the hidden variables $\lambda$. 
In the same way as the relatively simple laws of mechanics, when applied to a system with massively complex initial conditions, lead to the complexity of the laws of thermodynamics, Andreoletti \& Vervoort argue the complexity of quantum mechanics could arise from the simplicity of classical physical laws (or at least laws which don't require the violation of any of the other assumptions of e.g., Bell inequalities) combined with hugely complex initial conditions, represented through the dependence of the measurement settings in Bell-type experiments on the hidden variables.}
The issue of dependence on initial conditions is also addressed in literature by presenting certain superdeterminist toy models in which this dependence does not occur~\citep{tHooft2016cellular,DonadiHossenfelder2022,Ciepielewski2023,Hance2024}.

\subsection{The Objection of the Impossibility of Science}\label{subsec:ImpossOfSci}

This is one of the longest-standing objections to (\emph{na\"ive}) superdeterminist theories, often referred to as the Tobacco Company Syndrome~\citep{HossenfelderPalmer2020}. Fundamentally, it is an epistemological concern. In Maudlin's formulation~(\citeyear{Maudlin2019}), the objection draws an analogy with a tobacco industry apologist who first invokes a common cause to deny that smoking produces cancer, and then, when confronted with randomised animal trials, insists that the randomisation procedure itself was somehow biased in favour of placing cancer-prone subjects in the experimental group, a move Maudlin regards as wholly unscientific. It is not difficult to see the risks to scientific practice if this objection were to remain unresolved within superdeterminism. Such a denial of Statistical Independence would imply, for instance, the impossibility of discussing isolated systems or experimental random errors. Moreover, it would pose a significant risk to the empirical coherence of the theory, as defined by \cite{Barrett1996}.

According to Baas \& Le Bihan, in this case, the only admissible possibility would be that of an \textit{exceptionalist superdeterminism}, where SI is upheld as a valid premise solely in cases involving violations of Bell’s inequalities. In this sense, there would exist \textit{``some kind of cosmic principle, a fundamental law, or a hand of God, preventing us from using those systems to set the measurement settings''}~\citep[p.16]{Baas2023}.

The objection by Baas \& Le Bihan aims to critique the \textit{ad hoc} nature of superdeterminism, but it has a straightforward response within such a framework---NSD, in fact, claims to provide precisely that ``kind of cosmic principle'' which the authors find objectionable:
\begin{displayquote}
    \textit{``[…] the dependence relation she [the superdeterminist] contemplates involves a highly specific class of variables $\lambda$, not just the “usual” physical properties […] these variables are part of a theory that describes the Big Bang (and everything after it), so a (still elusive) theory of quantum gravity, or rather a (still more elusive) “Theory of Everything” (ToE).''}~\citep[p.3]{Andreoletti2022}
\end{displayquote}

Na\"ive superdeterminism, therefore, deflects the accusation of \textit{ad hoc-ness} or exceptionalism by referring to a more fundamental explanatory framework capable of justifying the apparent ``hand of God'' behind the violation of SI. The solidity of SI in non-quantum phenomena is then justified by referring to its approximate nature: the correlation terms would, in a sense, have a fragility that makes them undetectable at scales higher than the fundamental one. This argument is similar to other less exotic cases in quantum mechanics, such as decoherence~\citep{Zurek1991Decoherence,Bacciagaluppi2025Decoherence}.

\subsection{The Weight of the Fundamental}\label{subsec:Weight}

As mentioned above, there are other arguments against superdeterminism, such as the objections from implausibility~\citep{Chen2021,Baas2023}, initial conditions~\citep{HossenfelderPalmer2020,Chen2021,Andreoletti2022,Baas2023}, or fine-tuning~\citep{HossenfelderPalmer2020,Chen2021,Ciepielewski2023}. However, in all these cases, the same principle applies as in the two previous examples: to defend itself from these objections, \textit{na\"ive superdeterminism} must necessarily present itself as a fundamental theory. 

This point has been explicitly emphasised on several occasions by proponents of superdeterminism. \cite{Andreoletti2022}, for example, explicitly refer to a more fundamental level precisely when it comes to refuting objections regarding the impossibility of science and, consequently, the difficulties associated with violating statistical independence---see:
\begin{displayquote}
    \textit{This argument has more traction, we believe, if one realises that the $\lambda$ are truly unique, in the most absolute sense: they are the variables occurring in the ultimate ToE – not the mundane variables we encounter every day, for which statistical independence remains obvious. Note that the above argument still holds \textbf{even} if physicists would never be able to construct such a ToE in practice.}~\citep[p.23]{Andreoletti2022}
\end{displayquote}
and
\begin{displayquote}
    \textit{Superdeterminism boils down to a \textbf{theoretical}, \textbf{in-principle} dependence (on the ab-initio variables of the ultimate ToE).}~\citep[p.25]{Andreoletti2022}
\end{displayquote}
Similarly, \cite[p.3]{Hossenfelder2020Superdeterminism} also explicitly refers to such a more fundamental level for this reason.
In this claim to fundamentality, however, NSD takes on the characteristics of a metaphysical theory rather than a scientific one.

It is important to clarify what is meant by \textit{fundamental} here. The topic is clearly broad~\citep{Tahko2023Fundamentality}, and our brief discussion does not aim to exhaust it. Specifically, we want to clarify what is meant by \textit{fundamental} in physics. We will then define both strong and weak forms in which we can speak of a fundamental theory in a maximal sense:

\begin{definition}\label{Def:SFPT}[\textbf{Strongly Fundamental Physical Theory (SF)}] A physical theory is fundamental in a strong sense iff (if and only if) it satisfies the following conditions~\citep[p.6]{Crowther2019} and \citep[p.3]{Morganti2020b}:
\begin{itemize}\label{CONDITIONS}
    \item Predictive at all scales;
    \item Non-perturbative;
    \item Natural—i.e., avoiding fine-tuning;
    \item Background-independent— no \textit{ad-hocness};
    \item Internally unified;
    \item Unique;
    \item Internally consistent;
    \item Definite.
\end{itemize}
\end{definition}
This set of conditions implies that a strongly fundamental theory cannot be reduced to other, less fundamental theories. This is as a strongly fundamental theory, being necessarily both unique and predictive at every scale, cannot admit the existence of scales for which it does not predict, or competing theories in its own domain of application and thus any theory to which it could potentially be reducible. Reduction will therefore always occur asymmetrically from non-fundamental theories to the fundamental one.

\begin{definition}\label{WFPT}[\textbf{Weakly Fundamental Physical Theory (WF)}] \citep{Cao2003,Morganti2020a} A physical theory is fundamental in the weaker sense: \textit{``if it accounts for the possibility of using specific subsets of the criteria identified for the stronger sense, in a relative and contextual sense, such that one theory is considered more fundamental than another''}~\citep[p.4]{Morganti2020a}.
\end{definition}

According to \cite{Andreoletti2022} \textit{na\"ive superdeterminism} is admissible only by assuming the existence of a Strongly Fundamental Theory, thus excluding \textit{a priori} the possibility of non-fundamental metaphysical accounts such as Anderson’s non-layered conception (\citeyear{Anderson1972}), or infinitist theses (such as \cite{Dehmelt1989}) or coherentist views~\citep{Calosi2016}. As in \cite{Baas2023}, admitting \textit{na\"ive superdeterminism} entails the elimination of alternative metaphysical positions.

This, however, is not necessarily an issue. The fact that superdeterminism excludes from metaphysical reflection positions alternative to strong fundamentality, if superdeterminism had good reasons to be considered reliable, would not be regarded as a (major) problem. And if it were so, it would be a problem common to any other theory that presents itself as a strongly fundamental one. Trivially, NSD can only present itself as strongly fundamental (or at least weakly fundamental with respect to the most fundamental theory currently available) due to its violation of SI. Statistical independence is in fact {an assumption made by every purported fundamental physical theory in circulation}; NSD, contextualising this violation on a more fundamental level, simply interprets SD as the most fundamental theory with respect to those assuming SI. A similar argument can be found in \cite{Baas2023} when they discuss \textit{exceptionalist superdeterminism} (see Section~\ref{subsec:ImpossOfSci}).

The error of \textit{na\"ive superdeterminism} lies, more than anything, in how it uses strong fundamentality. Rather than identifying whether NSD fits the criteria of fundamentality given, its proponents instead assume strong fundamentality, using it as a premise for arguments aimed at legitimating \textit{na\"ive superdeterminism}.

One may wonder why recourse to fundamentality carries such justificatory power with respect to \textit{ad hoc} features of a theory. The answer is that \textit{ad hoc}-ness \textit{per se} is not undesirable when it can be recontextualised within a more comprehensive theoretical framework. Consider the following scenario: a chemist performing an exothermic reaction consistently records a temperature fractionally above the theoretical prediction, yet when the experiment is conducted remotely via robotic equipment, the recorded temperature matches the forecast exactly. This warrants the following law: ``Every time a human performs the experiment, reaction $X$ produces sufficient heat to raise the room temperature to $\Delta+p$ degrees, where $p$ denotes the additional increase and $\Delta$ the theoretically predicted value.'' Such a law is manifestly \textit{ad hoc}, yet its peculiarity is readily explained by embedding the scenario within a larger explanatory framework: the body heat released by the experimenter during preparation accounts for the recorded discrepancy. Contextual \textit{ad hoc-ness} is therefore not only unproblematic but often scientifically useful. However, this embedding procedure would be unavailable in the case of a fundamental theory; seeking an explanation for such \textit{ad hoc-ness} would amount to positing a further link in an explanatory chain that, by stipulation, admits of no additional theoretical constraint.

Therefore to support \textit{na\"ive superdeterminism} is to support an argument like the following:
\begin{enumerate}[label=\alph*).]
    \item\label{NaiveSuperdetArgA} There exists a fundamental physical theory;
    \item\label{NaiveSuperdetArgB} Superdeterminism is the fundamental physical theory;
    \item\label{NaiveSuperdetArgC} Since superdeterminism is the fundamental physical theory, we can justify certain peculiarities about it.
\end{enumerate}

This argument is clearly illegitimate. In order to assert \ref{NaiveSuperdetArgB}, NSD should provide evidence of complying with at least the criteria that we have indicated as characteristic of fundamentality (i.e., those in Def.~\ref{Def:SFPT}). Moreover, as we have seen in \ref{CONDITIONS}, a theory being background-independent---i.e. not \textit{ad hoc}---is in fact one of the criteria given by \cite{Crowther2019} for that theory being strongly fundamental, contrasting with the argument of the \textit{ad hoc}-ness of superdeterminism, proposed by Baas \& Le Bihan. However, none of the na\"ive superdeterminism proposals explicitly endorse the argument above. 

\noindent
Essentially, instead of stating
\begin{enumerate}
    \item Superdeterminism is not \textit{ad hoc};
    \item Avoiding \textit{ad hoc}-ness is a feature of a (strongly) fundamental theory;
    \item Thus superdeterminism has a feature of a (strongly) fundamental theory.
\end{enumerate}
Androletti and Vervoort state that
\begin{enumerate}\label{SUPERDET}
    \item Superdeterminism is (part of) a fundamental theory;
    \item \textit{Ad hoc}-ness must be avoided by any fundamental theory;
    \item Therefore, the accusation of superdeterminism being \textit{ad hoc} cannot be made.
\end{enumerate}
But, regardless of this reformulation, without offering any further model or evidence for this fundamental theory, the argument remains unacceptable. In fact, the erroneous idea of deriving truthfulness of a theory in virtue of its fundamental status is far from being an exclusive presumption of superdeterminism. 

\subsection{A Formal Flaw}\label{AFORMALFLAW}
One point that should not be overlooked from the previous paragraph is the formal nature of our critique of superdeterminism. We are not denying, in other words, the possibility that:
\begin{enumerate}[label=\alph*.] 
    \item  \label{FUNDAMENTALTHEORYEXISTENT} There exists a fundamental physical theory;
    \item There exists a corroborated theory $X$ that reflects the characteristics of strong fundamentality (i.e. some form of superdeterminism);
    \item $X$ (i.e. some form of superdeterminism) is a fundamental physical theory.
\end{enumerate}

However, the crucial feature of such arguments, is that the strong fundamentality claim for $X$ takes place in virtue of the presence of other empirical or meta-empirical advantages possessed by a given theory, whose fundamental status is granted only \textit{ex post}, when these same advantages have been assessed~\citep{Mcmullin1982,Kuhn2024}. 
Thus, the previous conclusion of ours should not be mistaken for a critique of the notion of fundamentality in physics \textit{per se} or as a statement that the identification of strongly fundamental physical properties has no value for the scientific enterprise. On the contrary, metaphysical considerations on the fundamental status of certain physical theories have been proven to be extremely useful on various occasions in the history of physical sciences.

Classical physics abounds with such cases; just consider, for instance, the development of the concept of inertial mass as a fundamental notion in mechanics, wholly separate from the related concept of \textit{quantity of matter}. Inspired by Newton's work, the treatment of quantity of matter as synonymous for physical mass can be found in the writings of British physicists such as \cite{Maxwell1991} and \cite{Thomson2022} as late as the end of the 19th century. This definition made the inertial behaviour of bodies wholly dependent on the conjunction of other supposedly more primitive properties such as density and volume, without granting a fundamental status to inertial mass as such. However, on the continent, thanks to the work of authors such as Leibniz and Euler, the idea of inertia as a fundamental property of bodies gradually emerged~\citep{Jammer1997}. Building on these developments, Mach was able to propose an operational definition of mass, completely separate from the recourse to statements concerning the internal structure of bodies, and establishing the measure of inertia as \textit{the} fundamental conception of mass~\citep{Koslow1968}.

But what makes the case of the theory of mass so different from the one presented by superdeterministic theories? The two situations make recourse to fundamentality in opposite ways. Mach argued in favour of his inertial theory on the basis of the evident epistemic advantage of talking about mass in terms of inertia. This is because while we cannot directly access the densities of different bodies, or describe their dynamics with reference to volume only, we have immediate empirical access to their inertial properties through the measurements of their accelerations. In the case of superdeterminism, the situation is reversed. There is no advantage under an epistemological point of view in supporting a variant of superdeterminism (as we have shown in the case of NSD theories) and fundamentality must come into play to rescue the theory from its own epistemological shortcomings. In this way, fundamentality in \textit{na\"ive superdeterminism} is just postulated as a precondition for the validity of the theory, instead of being granted by virtue of the empirically-verifiable satisfaction of the requirements of Def.~\ref{Def:SFPT}.

\textit{Na\"ive superdeterminism} places a metaphysical demand at the forefront, trying, based on this, to legitimise itself as a scientifically acceptable theory. This \textit{petitio principi} can only result in the shift superdeterminism from the domain of scientific proposals to that of metaphysical ones.
{In this respect, the appeal to fundamentality alone is not enough to guarantee the pursuitworthiness of the superdeterminist project. As has recently been emphasised by a growing literature on the topic {\citep{douglas2013, Lichtenstein2021, duerrfischer}}, pursuitworthiness crucially depends on a research program’s ability to exhibit a range of theoretical virtues: simplicity, accuracy, etc. 
In the case of other well-known research programs that are currently being pursued in theoretical physics, such as Loop Quantum Gravity and String Theory, their pursuitworthiness is usually defended in terms of their unificatory power. 
Superdeterminism as a research program however, comparatively lacks a similar ``rationally warranted promise'' to deliver a payoff, beyond its endorsement of locality and determinism. As we have argued, the only potential exception in this regard, its alleged fundamentality, is insufficient to argue for its pursuitworthiness by itself.}

Additionally, this postulation of fundamentality carries other serious and unwarranted consequences for superdeterminists.
For example, returning to Crowther's list (Def.~\ref{Def:SFPT}), the uniqueness constraint implies that following \ref{SUPERDET}, being superdeterministic would turn out to be a requirement for fundamentality as well. In this context, other fundamental research programs---such as loop quantum gravity~\citep{Rovelli2014}---would essentially be discarded from the rank of potentially \textit{fundamental} theories, for no reason other than that they are not superdeterministic. 
{This does not mean that superdeterminists cannot explore their proposals, but rather that any defence of the fundamentality of superdeterminism must confront the challenge of demonstrating its undisputed superiority over potential competitors. Such a defence could in principle be structured along the lines of a No Alternatives Argument, similarly to what has been proposed in the context of string theory, which has already faced a similar difficulty~{\citep{Dawid2015}}. However, the superdeterminist framework currently falls far short of being a realistic candidate for the only viable path toward a ToE. Unlike string theory, superdeterminism has yet to clearly establish what its concrete payoffs would be, nor how these compare against alternative research programmes.}

We arrive then at an apparent paradox: a simple formal flaw forces us to reject a theory that could, \textit{de facto}, prove to be correct. This is not actually a new phenomenon: what makes science science is, so to speak, a matter of method.

\section{Metaphysical Superdeterminism}\label{sec:MetaSoln}

Having shown that the type of superdeterminism represented by \textit{na\"ive superdeterminism} is characterised more as a \textit{metaphysical} presupposition than a scientific one, we now turn our the analysis to \textit{metaphysical superdeterministic} proposals. While in Sec.~\ref{sec:Metaphys} we identified an \textit{epistemological} problem with NSD models (i.e., that it claimed fundamentality without presenting a theory which we otherwise would have reasonable grounds to consider fundamental), the same argument cannot be used against MSD models. This is because MSD models explicitly set out to develop the sort of \textit{fundamental theory} which NSD models simply assert they are part of (to justify themselves against criticism). The problem with MSD is therefore not an improper claim to \textit{fundamentality}, but rather (as we will argue) the construction of a confused and unpalatable metaphysical framework.

While some proposals for MSD (such as Leibnizian Quantum Mechanics) have already had specifically metaphysical objections developed against them~\citep{Chen2021,Ciepielewski2023}, the majority of MSD positions are still attacked by appealing to epistemic criteria.

While the conclusions of Sec.~\ref{sec:Metaphys} held in general for any NSD, it is clear that, as specific physical theories, for MSD it is necessary to develop arguments that concern each specific proposal. However, in order to show the difficulties that superdeterminism faces when trying to build such \textit{fundamental theories}, we'll consider the most robust form of MSD currently available---supermeasured theories~\citep{HanceHossenfelderPalmer2022,Hance2024,Palmer2024}. By Wharton and Argaman's definitions, supermeasured theories form a subclass of superdeterministic theories, being as they treat the measurement settings $a$ and $b$ as statistical parameters and violate the no-conspiracies assumption.

The core idea behind supermeasured theories is quite simple: since violating SI implies a level of conspiracy and fine-tuning that we find unacceptable, supermeasured theories propose deriving the SI violation from a non-trivial measure on the state space. This approach avoids the unappealing causal dependence of the (set of) measurement choices $a$ on the hidden variables $\lambda$, referring instead to (a set of) measurement choices $a$ that are consistent with particular measures; this is because certain combinations of $a$ and $\lambda$ end up having measure zero, thereby imposing a restriction on counterfactuals. This allows us to reject non-locality (in the Einsteinian sense) without having to rely on any kind of piecemeal fine-tuning between hidden variables and detector settings.

In other words, let’s consider again SI (Eq.~\eqref{eq:SI}):
\begin{center}
    $\rho(\lambda|a)=\rho(\lambda)$
\end{center}     
This condition asserts the equivalence between two probability distributions, normalised over a space we will call $\mathcal{S}_{math}$, which represents the space of all mathematically possible states. However, when we integrate over a space like $\mathcal{S}_{math}$, we require a certain measure $\mu(\lambda,a)$. This implies that the quantity actually involved in the derivation of Bell inequalities is not $\rho$ on $\mathcal{S}_{math}$, but rather:
\begin{equation}
    \rho_{Bell}=\rho(\lambda,a)\mu(\lambda,a)
\end{equation}
In the case of a trivial measure, it naturally follows that $\rho_{Bell}=\rho$. Thus, SI as it truly applies in Bell inequalities becomes:
\begin{equation}\label{eq:BellSI}
    \rho_{Bell}(\lambda|a)=\rho_{Bell}(\lambda)
\end{equation}
A supermeasured theory, then, is a theory that violates Eq.~\eqref{eq:BellSI} without violating Eq.~\eqref{eq:SI}, by proposing an alternative measure $\mu*$ instead of the typically assumed measure $\mathcal{\mu}$, which is the uniform measure over $\mathcal{S}_{math}$ (since in Bell’s derivations, this measure is never specified). This non-trivial measure $\mu$ is known as a \textit{supermeasure}, which is the origin of the name for these theories.

Currently, there are several examples of non-trivial measures for supermeasured theories, such as IST~\citep{Palmer2009,Palmer2014,Palmer2016,HanceHossenfelderPalmer2022} and Rational Quantum Mechanics (RaQM)~\citep{Palmer2024}. For additional discussion of these measures, see \cite{Hance2024,HanceHossenfelderPalmer2022}. We will focus on IST, as it is the most thoroughly-developed option available.

\subsection{Invariant Set Theory}\label{subsec:IST}

IST is presented as a
\begin{displayquote}
    \textit{``realistic, locally causal theory of fundamental physics which assumes a much stronger synergy between cosmology and quantum physics than exists in contemporary theory. In IST the (quasi-cyclic) universe $\mathcal{U}$ is treated as a deterministic dynamical system evolving precisely on a measure-zero fractal invariant subset $\mathcal{I}_u$ of its state space. In this approach, the geometry of $\mathcal{I}_u$, and not a set of differential evolution equations in space-time $\mathcal{M}_u$, provides the most primitive description of the laws of physics.''}~\citep[p.1]{Palmer2016}
\end{displayquote}

IST, by treating the geometry of $\mathcal{I}_u$ as the most primitive expression of physical laws---rather than the set of differential equations described above---makes a conceptually nuanced move. Indeed, accepting $\mathcal{I}_u$ as fundamental (rather than anything inherently spatio-temporal) implies the presence of top-down constraints (of cosmological origin), which stand in stark contrast to the bottom-up constraints typical of the standard reductionist approach it critiques. We will see later in Sec.~\ref{sec:Ambig} how this has profound consequences for the metaphysics of IST.

For now, this is enough for us to conclude that:
\begin{proposition}\label{FUNDAMENTALSTRUCTURE}
   \textbf{The fundamental structure of reality described by IST is not spatio-temporal in a relativistic sense.}
\end{proposition}
\noindent 
as is repeatedly emphasised by \cite{Palmer2014,Palmer2016}. We will now look at how IST has been concretely helpful in the context of Bell’s theorem, by providing a toy example of a supermeasured theory in action~\citep{HanceHossenfelderPalmer2022}.

\subsection{Violating SI without violating SI}\label{subsec:Supermeasd}

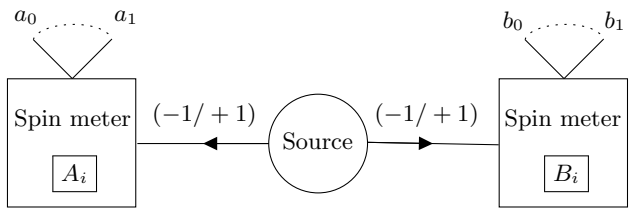
\begin{figure}
    \centering
\begin{tikzpicture}[x=0.75pt,y=0.75pt,yscale=-0.8,xscale=0.8]

\draw   (300,93.14) .. controls (300,75.94) and (313.94,62) .. (331.14,62) .. controls (348.34,62) and (362.28,75.94) .. (362.28,93.14) .. controls (362.28,110.34) and (348.34,124.28) .. (331.14,124.28) .. controls (313.94,124.28) and (300,110.34) .. (300,93.14) -- cycle ;
\draw    (217.17,93.14) -- (300,93.14) ;
\draw    (261.58,93.14) -- (300,93.14) ;
\draw [shift={(258.58,93.14)}, rotate = 0] [fill={rgb, 255:red, 0; green, 0; blue, 0 }  ][line width=0.08]  [draw opacity=0] (8.93,-4.29) -- (0,0) -- (8.93,4.29) -- cycle    ;
\draw    (444.99,93.92) -- (362.17,92.37) ;
\draw    (400.58,93.09) -- (362.17,92.37) ;
\draw [shift={(403.58,93.14)}, rotate = 181.07] [fill={rgb, 255:red, 0; green, 0; blue, 0 }  ][line width=0.08]  [draw opacity=0] (8.93,-4.29) -- (0,0) -- (8.93,4.29) -- cycle    ;
\draw   (135.28,52.28) -- (216.17,52.28) -- (216.17,133.17) -- (135.28,133.17) -- cycle ;
\draw    (151.6,27.6) -- (176.17,52.17) ;
\draw    (176.17,52.17) -- (201.17,27.17) ;
\draw    (461.6,27.6) -- (486.17,52.17) ;
\draw    (486.17,52.17) -- (511.17,27.17) ;
\draw  [dash pattern={on 0.84pt off 2.51pt}]  (151.6,27.6) .. controls (155.28,15.85) and (195.28,13.85) .. (201.17,27.17) ;
\draw  [dash pattern={on 0.84pt off 2.51pt}]  (461.6,27.6) .. controls (465.28,15.85) and (505.28,13.85) .. (511.17,27.17) ;
\draw   (445.28,52.28) -- (526.17,52.28) -- (526.17,133.17) -- (445.28,133.17) -- cycle ;

\draw (307,85) node [anchor=north west][inner sep=0.75pt]   [align=center] {$\displaystyle \text{Source}$};
\draw (365,66) node [anchor=north west][inner sep=0.75pt]   [align=left] {$\displaystyle (-1/+1)$};
\draw (225,66) node [anchor=north west][inner sep=0.75pt]   [align=left] {($\displaystyle -1/+1)$};
\draw (162,99) node [anchor=north west][inner sep=0.75pt]   [align=left] {$\boxed{\displaystyle{A_i}}$};
\draw (472,99) node [anchor=north west][inner sep=0.75pt]   [align=left] {$\boxed{\displaystyle B_i}$};
\draw (138,8) node [anchor=north west][inner sep=0.75pt]   [align=left] {$\displaystyle a_{0}$};
\draw (201,8) node [anchor=north west][inner sep=0.75pt]   [align=left] {$\displaystyle a_{1}$};
\draw (446,8) node [anchor=north west][inner sep=0.75pt]   [align=left] {$\displaystyle b_{0}$};
\draw (510,8) node [anchor=north west][inner sep=0.75pt]   [align=left] {$\displaystyle b_{1}$};
\draw (138,70) node [anchor=north west][inner sep=0.75pt]   [align=left] {Spin meter};
\draw (447,70) node [anchor=north west][inner sep=0.75pt]   [align=left] {Spin meter};

\end{tikzpicture}

\caption{Simplified diagram of the CHSH experiment \citep{ClauserShimonyHolt1969}.
A source emits pairs of spin-entangled particles in opposite directions. The left-hand particle (A) is subjected to either test $a_0$ or $a_1$, while the right-hand particle (B), is subjected to either test $b_0$ or $b_1$, with each test having $+1/-1$ as possible results.}
    \label{fig:CHSH}
\end{figure}

Let us imagine a scenario similar to that of \cite{ClauserShimonyHolt1969} (see Fig.~\ref{fig:CHSH}) and call $\Lambda$ the space of all hidden variables $\lambda$, dividing it into subsets characterised by each combination of \textit{result} and \textit{setting}, $\Lambda^{AB}_{XY}$. This means that, for example, the subset $\Lambda^{++}_{00}$ contains all the $\lambda$'s for which, for the setting $X_0 Y_0$, the outcomes $A = +1$ and $B = +1$ will occur, and so on. Subsequently, we assume that the same $\lambda$ cannot occur for two different combinations of experimental results, i.e., the subsets of the form $\Lambda^{ij}_{kl}$ are mutually exclusive. In this way, some combinations of $\lambda$ and $(X,Y)$ are in $\mathcal{S}_{math}$, but are not physically possible.

Then, \textit{Supermeasured} theories make use of a measure that is not uniform over all $\Lambda$: a certain $\mu*$ must be introduced, which assigns a value 0 to certain combinations of $(X,Y,\lambda)$. 

In the case of IST, a rationality criterion is introduced: if the first combination of settings, $(X_0Y_0)$ for a given $\lambda$, satisfies this rationality condition, then the second, $(X_0 Y_1)$ for the same $\lambda$, cannot satisfy it---and so are associated with a state in $\mathcal{S}_{math}$ where $\mu = 0$:
\begin{equation}
\begin{split}
&\rho_{Bell}(\lambda|X_0Y_0) \not=0 \rightarrow \\
&\rho_{Bell}(\lambda|X_0Y_1) = \rho_{Bell}(\lambda|X_1 Y_0) = 0
\end{split}
\end{equation}
\begin{equation}
\begin{split}
&\rho_{Bell}(\lambda|X_1 Y_0) \not=0 \rightarrow\\
&\rho_{Bell}(\lambda|X_0Y_0) = \rho_{Bell}(\lambda|X_1 Y_1) = 0
\end{split}
\end{equation}
However, this does not imply that:
\begin{equation}
\rho(\lambda|X_0Y_0) \not=0 \rightarrow \rho(\lambda|X_0Y_1) = \rho(\lambda|X_1Y_0) = 0
\end{equation}
or
\begin{equation}
\rho(\lambda|X_1Y_0) \not=0 \rightarrow \rho(\lambda|X_0Y_0) = \rho(\lambda|X_1Y_1) = 0
\end{equation}

Thus, IST violates Eq.~\eqref{eq:BellSI} without violating Eq.~\eqref{eq:SI}.

\section{The Ambiguity of the Fundamental}\label{sec:Ambig}

In this section, we will demonstrate how IST is metaphysically ambiguous, specifically examining how this ambiguity centres on what IST assumes as fundamental entities. Our aim is twofold. On the one hand, we will show how the holism implicit in IST supports a form of priority monism~\citep{Schaffer2010Monism}; on the other, we will outline the metaphysical challenges this causes for IST.

\subsection{Monism for Superdeterminists}\label{subsec:Monism}

In order to reconstruct the metaphysical underpinnings of superdeterminism, we introduce a well-known formalism~\citep{Schaffer2010Monism,IsmaelSchaffer2020}. Let $D$ represent the relation of metaphysical dependence, $P$ the parthood relation, and $u$ the cosmos, understood as the maximal concrete whole~\citep{Schaffer2024Monism}.
\begin{eqnarray}
{D}_{xy} &\coloneqq x \text{ depends on } y,\\
P_{xy} &\coloneqq x \text{ is a part of } y,\\
u &\coloneqq \text{ the cosmos.}
\end{eqnarray}
Using this framework, we can easily define a concrete object $C_x$ as a part of the cosmos:
\begin{equation}
  C_x \coloneqq P_{xu}
\end{equation}
With $C_x$ defined, we can then derive a definition for $B_x$, the basic objects, which are concrete objects that do not depend on any other concrete objects:
\begin{equation}
B_x \coloneqq C_x \land \neg \exists y (C_y \land D_{xy})
\end{equation}
Thus, answering the question of what constitutes the fundamental mereology---i.e. identifying the basic concrete objects within a theory---boils down to identifying the entities $x$ in a theory $T$ that satisfy $B_x$.

Let us now introduce two minimal constraints on the properties required of our candidate $B_x$, such that the set of basic objects grounds the entirety of $u$ without gaps or overlaps:
\begin{equation}\label{eq:covering}
    \text{Covering: }\sum x (B_x) = u
\end{equation}
\begin{equation}\label{eq:nopart}
    \text{No Parthood: } \forall x \forall y (B_x \land B_y \land x \neq y) \Rightarrow \neg (P_{xy})
\end{equation}
That is, Eq.~\eqref{eq:covering} states that the sum of all entities $x$ grounded by a basis $B_x$ exhausts without exception the totality of all physical objects in our cosmos. In other words, the set of $B$ must be complete: duplicating all these entities (the set of $B_x$), while preserving their fundamental relations, should metaphysically suffice to duplicate the cosmos and its contents. This constraint is also known as the \textit{completeness requirement} \citep{Schaffer2010Monism}.

Eq.~\eqref{eq:nopart} instead states that the basic objects cannot be in a parthood relation with each other. 

At this point, we can define priority monism as:

\begin{definition}[Priority Monism]\label{def:PriorityMon}
The ontology of a model is Priority Monist iff $\exists x$ such that $B_x \land (\forall y,\,B_y \Rightarrow x = y)$
\end{definition}

And, more relevant for our purposes, a version that identifies the unique basic object with the whole:

\begin{definition}[Priority Monism (cosmic)]\label{def:CosPriorityMon}
    The ontology of a model is Cosmically Priority Monist iff $\exists ! x$ such that $B_x \land B_u$
\end{definition}

For completeness, we also note the definition of priority pluralism: 

\begin{definition}[Priority Pluralism]\label{def:PriorityPlu}
The ontology of a model is Priority Pluralist iff $\exists x \exists y$ such that $B_x \land B_y \land x \neq y$
\end{definition}

And the more familiar version, which associates the multiple basic objects with proper parts of $u$: 

\begin{definition}[Priority Pluralism (partial)]\label{def:PartPriorityPlu}
The ontology of a model is Partially Priority Pluralist iff $\exists x \exists y$ and $(B_x \land B_y \land x \neq y) \land \neg B_u$
\end{definition}

In conclusion to our formal repertoire we add an assumption ~\citep{Schaffer2010Monism} around the relationship between monism and pluralism that will prove useful to us later:
\begin{proposition}\label{GENERALMETAPHYSICAL}
      \textbf{Monism and pluralism are general metaphysical theories and, therefore, if true, necessarily true.}
\end{proposition}
The exclusivity of monism and pluralism is closely linked to Eq.s~\ref{eq:covering} and \ref{eq:nopart}. If we were to consider the whole spectrum of doctrines around fundamentals, we would also need to consider priority nihilism~\citep{Schaffer2024Monism}, i.e.$\neg\exists Bx$. However, given superdeterminism as presented above requires some notion of fundamentality, as discussed in Sec.~\ref{subsec:Weight}, and consequently some notion of priority, it is inherently incompatible with priority nihilism.

\subsection{Superdeterminism for Monists}\label{subsec:SuperdetForMonists}

Now, we will show how IST is holistic in a way that supports  priority monism as described above. There should be little doubt by now that IST supports a certain type of holism:
\begin{displayquote}
\textit{``The type of holistic primal geometry $\mathcal{I_U}$
discussed in this paper provides an example of the notion of a top-down influence, where some geometric constraint on the state space of the whole universe determines the properties of individual particles. This is in complete contrast to the more common bottom-up philosophy of reductionism.''}~\citep[p.15]{Palmer2019}
\end{displayquote}
The existence of top-down geometric constraints on the state space of the entire universe, as seen in Sec.~\ref{sec:Metaphys}, further attests sufficiently to the holistic nature of IST. We can therefore conclude that:
\begin{proposition}\label{ISTHOLISM}
    \textbf{IST supports a certain type of holism.}
\end{proposition}
But how does this holism support priority monism within IST?

To answer this question, we must first determine which type of holism is compatible with IST, given that the correlations established by the theory admit of two distinct types of holism. The first~\citep{Morganti2009RelationalHolism,Teller1986RelationalHolism,Teller1989RelativityHolismBell}, known as \textit{nomic} or \textit{correlational} holism, is the type generally found in discussions of Bell's theorem, where variables across spacelike-separated regions exhibit covariation that cannot be decomposed into local contributions. This type of holism is compatible with a broader range of metaphysical positions, including non-monistic ones, such as priority pluralism enriched with irreducible external relations. The second~\citep{Schaffer2010Monism,Schaffer2024Monism}, known as \textit{grounding} holism, is a stronger form and the one Schaffer has in mind, compatible only with priority monism. It can therefore be concluded that IST supports priority monism only if it supports grounding holism. That IST requires more than correlational holism is confirmed by Palmer's own characterisation of the model:

\begin{displayquote}
    \textit{The Invariant Set Postulate provides support to this search: treating the geometry of the invariant set as primitive, introduces a fundamentally atemporal perspective into the formulation of basic physics. Such a perspective is absent in classical physics, in which differential equations are considered primitive.}~\citep[p.10]{Palmer2009}
\end{displayquote}
and similarly in \cite[p.8]{Palmer2014}.
In any case, regardless of Palmer’s personal views on the nature of IST, it would be highly problematic for IST to endorse a form of correlational holism because it would reintroduce precisely the kind of correlation that the theory itself aims to eliminate, merely re-described via the supermeasure. This is why Palmer must admit (or rather, state boldly) that correlations are due to the ontologically fundamental laws of the universe’s evolution within its fractal subspace. We can therefore conclude that
\begin{proposition}\label{KINDOFHOLISM}
    {\textbf{IST supports only grounding holism}}
\end{proposition}
{To demonstrate that IST supports a form of priority monism, we now must simply show that} there is no pluralistic basis that satisfies Eq.~\eqref{eq:covering} for IST\footnote{This demonstration follows the same one provided by \cite{Schaffer2010Monism}.}.

Consider the classic Bell-type experiment in an IST setting. Suppose, as before, that there exists a triplet $(X,Y,\lambda)$ where only certain combinations of $X,Y$ and $\lambda$ are permitted via $\mu$. We define the property, for a triplet $(X,Y,\lambda)$, of having a non-zero $\rho_{Bell}$  ($\mathcal{F}$) and we consider a Democritean pluralist base~\citep{Schaffer2010Monism}---a pluralist base where the basic objects are mereologically minimal.

We now wonder whether there is any Democritean base that can fix the property $\mathcal{F}$ for the triplet $(X,Y,\lambda)$. But given proposition \ref{ISTHOLISM}, we know that property $\mathcal{F}$ does not derive from bottom-up constraints. Consequently, no Democritean base satisfying Eq.~\eqref{eq:covering} for its $x$ such that $B_x$\ can account for $\mathcal{F}$, meaning the completeness requirement fails\footnote{We will not discuss the possibility of emerging properties here, and we refer to the discussion addressing the issue given by \cite{Schaffer2010Monism}.}.

Since this holds for a Democritean base, we can easily conclude that the same would apply to any pluralist base, as the properties of the whole do not appear to be determined by the total intrinsic properties of any subsystems---the situation should therefore not change considering non-minimum subsystems. This means we can conclude that
\begin{proposition}\label{NOPLURALISTIC}
    \textbf{There is no pluralistic basis that satisfies Eq.~\eqref{eq:covering} for IST}
\end{proposition}
Which in conjunction with Proposition~\ref{GENERALMETAPHYSICAL} allows us to conclude that:
\begin{proposition}\label{IST SUPPORT PRIORITY MONISM}
    \textbf{IST supports priority monism}
\end{proposition}

\subsection{An (Un)biased mereology?}

We will now discuss how this monistic nature of IST leads to certain metaphysical challenges. Note that in this section, in applying the concept of \textit{parthood relation} as a \textit{fundamentality device}---as in the example of \textit{substantivalist monism}---closely follows the approach proposed by \cite{LeBihan2018} for quantum gravity programs.

We know from the Sections~\ref{sec:Metaphys} and \ref{sec:MetaSoln} that the fundamental structure of reality described by IST is not spatio-temporal in a relativistic sense.

Taking proposition \ref{FUNDAMENTALSTRUCTURE}, definition \ref{def:CosPriorityMon} and the conclusion of \textit{supra} into account then, it follows easily for IST that:
\begin{proposition}\label{NONSPATIOTEMPORAL}
    \textbf{The maximal concrete whole for IST must be non spatio-temporal.}
\end{proposition}

But this creates a serious problem for the metaphysics of IST. To show how, following \cite{LeBihan2018}, let us consider briefly a different approach i.e. monistic substantivalism~\citep{Schaffer2009,Schaffer2010Monism}. In this account the cosmos, i.e., the maximal concrete whole, is identified with a substantivalist spacetime. Although this is not strictly required by any form of priority monism, monistic substantivalism succeeds in this way in using the parthood relation as a fundamentality device $D$~\citep{LeBihan2018}. As Le Bihan put it:
\begin{enumerate}[label=LB\roman*.]
    \item\label{LBi} Spacetime regions are proper parts of the whole spacetime.
    \item\label{LBii}  Proper parts are less fundamental than the whole they compose.
    \item\label{LBiii} Spacetime regions are derived from the whole they compose.
\end{enumerate}
In other terms, spacetime, or the cosmos, is fundamental (\ref{LBii}), and its spatio-temporal regions, as its parts (\ref{LBi}), are derived (\ref{LBiii}). Let us now return to IST: what is the maximal concrete whole for the theory? Taking Proposition.~\ref{IST SUPPORT PRIORITY MONISM} into account, we can argue that an intuitive approach might identify the cosmos with the universe $\mathcal{U}$ evolving on a measure-zero fractal invariant subset $\mathcal{I_U}$ of its state space. At this point, however, when we look at the metaphysical dependence relation between IST fundamental and derived structure, a metaphysical puzzle arises. Since the whole is fundamentally non-spatio-temporal \ref{NONSPATIOTEMPORAL}, the parthood relation becomes effectively unintelligible, end ends up failing as a viable fundamentality device.

Indeed, if we attempt to describe the relationship between the whole---as seen before---and spacetime (or any $x$ such that $C_x$ in IST) using parthood, we arrive at:
\begin{enumerate}[label=\alph*).]
    \item\label{ISTParthoodCdn:SpaceTime} Spacetime is a proper part of the whole.
    \item\label{ISTParthoodCdn:Part} Proper parts are less fundamental than the whole they compose.
    \item\label{ISTParthoodCdn:Deriv} Spacetime is derived from the whole it composes.
    \end{enumerate}
But conclusion \ref{ISTParthoodCdn:Deriv} cannot be earned in IST, since the terms \textit{proper part} and \textit{compose} in \ref{ISTParthoodCdn:SpaceTime} and \ref{ISTParthoodCdn:Part} cannot be understood except in a spatio-temporal setting.

Stating therefore that:
\begin{proposition}
    \textbf{The parthood relationship needs a spatio-temporal framework.}
\end{proposition}
From this and the previous conclusion, we see that:
\begin{proposition}
    \textbf{The parthood relation fails as a fundamentality device for IST}
\end{proposition}
IST would thus be a form of monism unable to account for the relations between part and whole, no longer being viable to conceive of parts as spatially defined.

In this sense, IST is comparable to loop quantum gravity (LQG)~\citep{Rovelli2014}, where the fundamental structure of reality is also non-spatio-temporal in relativistic terms. However, if the issue for LQG takes the form of an inquiry around the ontological status of relativistic spacetime---the well-known problem of \textit{spacetime emergence}~\citep{LeBihan2018,Huggett2018}---for IST the problem concerns the relations of fundamentality between its mereological structures.

To avoid all this IST therefore faces a profound mereological challenge: it must redefine the parthood relation $P$. Viable proposals already exist; \cite{LeBihan2018} suggests, for example, removing any spatial reference---explicit or intuitive---from the definition of parthood relation, developing a logical mereology, as proposed by \cite{Paul2002}. In this approach, the cosmos decomposes into spatio-temporal substructures understood as proper logical parts. Another possible route could be to follow the ``coherentist'' proposal by \cite{Morganti2021} for framing the ontology of QM in terms of structures of symmetric non-mereological dependence relations thus abandoning the usual layered conception endorsed by usual accounts of structuralism and holism. While the metaphysical picture presented by the two authors is concerned specifically with entanglement without referring to superdeterminism, the framework could potentially be extended to describe the metaphysics of IST.

However, these proposals, while promising, remain still underdeveloped and are neither considered nor discussed by proponents of IST. Indeed, the issue is entirely overlooked, leaving (this form of) superdeterminism, for now, in a state of profound metaphysical unpalatability.

\section{Discussion}\label{sec:Disc}

We have shown that, regardless of whether it takes the form of NSD or of MSD (at least for IST, the best MSD available so far), we have strong reason to regard superdeterminism with suspicion. In the case of NSD, this conclusion follows from the presence, within its theoretical structure, of claims of fundamentality that characterise it as a theory more metaphysical than scientific. In the case of IST, by contrast, the problem lies in the confused metaphysics it presupposes. {While criticisms of IST might not extend to other MSD proposals, or even future developments of IST itself, this problem means an appeal to IST in the context of a ``reappraisal'' of superdeterminism fails to provide a compelling case that the scepticism around superdeterminism is ill-motivated. Further, IST does not entail clear scientific promises for its endorsement beyond salvaging locality and determinism. Unlike programs such as Loop Quantum Gravity, IST does not yet offer compensating explanatory benefits that would offset the current challenge of providing a convincing account of spacetime emergence from the invariant set. While this doesn't stop one from investigating IST, it again means IST is still far from being a proposal capable of lending credence to the superdeterminism project over its detractors.} 

Returning to the claims of Sec.~\ref{sec:Whatare}, it thus seems that superdeterminism has no viable escape route (so far). A \textit{na\"ive superdeterministic theory} faces three possibilities: it may either fail to answer the objections; or it may present itself as fundamental---thus falling back into the issues discussed in Sec.~\ref{sec:Metaphys}; or it may construct an MSD capable of subsuming it. Yet, the one MSD proposal examined so far {remains} metaphysically obscure.

But what is the legitimacy of these arguments? In other words, are they sufficient to deny the \textit{de jure} possibility of a fundamental superdeterministic theory? Our view is that, while these arguments may not be sufficient \textit{de jure}, they are entirely legitimate, and this  formal insufficiency should not be of much comfort to superdeterminism.

{Note however that our argument does not imply that all research on superdeterminism ought to be abandoned. Within the context of what Shaw, drawing on Feyerabend, calls \textit{luxury science}~(\citeyear{Shaw2022})---that is, long-horizon research unconstrained by immediate empirical demands---superdeterminism can still find a legitimate place. However, in light of the considerations set out above, that place remains confined, for the time being, to the speculative domain - at least until the development of either a ``rationally warranted promise'' for the superdeterministic programme to deliver a payoff beyond locality and determinism, or a defence of the fundamentality of superdeterminism (along the lines of String Theory's No Alternatives Argument).
On this point, \cite{Fischer2026} has noted, regarding the pursuitworthiness of scientific experiments, that certain experiments may prove pursuitworthy insofar as they are capable of testing several competing theories within a single experimental instance. It is therefore not inconceivable that, even in the case of superdeterminism, favourable conjunctures could be exploited to collaterally test proposals such as that IST, as recently suggested by \cite{Hance_2025,palmer2025testingquantummechanicsquantum}.

In recent years, \cite{Ladyman2007,French2012,Maudlin2007} have warned against the unphysical presuppositions of armchair metaphysics, which, although successful in its own regard, tends to avoid serious involvement with cutting-edge physical knowledge. But if metaphysicians need to take more care of the physical implications of their theories, and the license they take when making physically dubious claims, surely so do physicists. As our analysis of NSD and MSD shows, even professional physicists may be altogether too quick in expressing theories whose deeper metaphysical implications are at risk of going unnoticed. This is especially true when attempting to shield one's hypotheses from undesired criticism by framing them such that they end up resembling purely metaphysical theories. Re-framing originally-physical theories in entirely metaphysically terms risks relinquishing the scientific rigour that should have been their primary objective.


\begin{bmhead}[Acknowledgements] 
GC and LM thank Lorenzo Bartalesi and Ferdinando Maieron for their invaluable advice during the early stages of writing of this work.
\end{bmhead}

\begin{bmhead}[Declarations of Competing Interests]
    None to declare.
\end{bmhead}

\begin{bmhead}[Funding Information]
JRH acknowledges support from a Royal Society Research Grant (RG/R1/251590), an EPSRC Mathematical Sciences Small Grant (UKRI3647), and their EPSRC Quantum Technologies Career Acceleration Fellowship (UKRI1217).
\end{bmhead}


\begin{thebibliography}{72}
\expandafter\ifx\csname natexlab\endcsname\relax\def\natexlab#1{#1}\fi

\bibitem[{{Adlam et~al.}(2024){Adlam, Hance, Hossenfelder, and Palmer}}]{adlam2023taxonomy}
{Adlam, E., Hance, J.~R., Hossenfelder, S. and Palmer, T.~N.} (2024) {Taxonomy for Physics Beyond Quantum Mechanics}. \textit{Proceedings of teh Royal Society A}. {\it 480}(20230779).
\newblock \doi{10.1098/rspa.2023.0779}.

\bibitem[{{Anderson}(1972)}]{Anderson1972}
{Anderson, P.~W.} (1972) More is different. \textit{Science}. {\it 177}(4047), 393--396.
\newblock \doi{10.1126/science.177.4047.393}.

\bibitem[{{Andreoletti and Vervoort}(2022)}]{Andreoletti2022}
{Andreoletti, G. and Vervoort, L.} (2022) Superdeterminism: a reappraisal. \textit{Synthese}. {\it 200}(5), 361.
\newblock \doi{10.1007/s11229-022-03832-6}.

\bibitem[{{Baas and Le~Bihan}(2023)}]{Baas2023}
{Baas, A. and Le~Bihan, B.} (2023) What does the world look like according to superdeterminism? \textit{The British Journal for the Philosophy of Science}. {\it 74}(3), 555--572.
\newblock \doi{10.1086/714815}.

\bibitem[{{Bacciagaluppi}(2025)}]{Bacciagaluppi2025Decoherence}
{Bacciagaluppi, G.} (2025) The role of decoherence in quantum mechanics In {\em The Stanford Encyclopedia of Philosophy},  Zalta, E.~N. and Nodelman, U. (eds). Metaphysics Research Lab, Stanford University. spring 2025 edition.
\newblock Available at: \url{https://plato.stanford.edu/archives/spr2025/entries/qm-decoherence/}.

\bibitem[{{Barrett}(1996)}]{Barrett1996}
{Barrett, J.~A.} (1996) Empirical adequacy and the availability of reliable records in quantum mechanics. \textit{Philosophy of Science}. {\it 63}, 49–64.
\newblock \doi{10.1086/289893}.

\bibitem[{{Bell}(1964)}]{Bell1964}
{Bell, J.~S.} (1964) {On the {E}instein {P}odolsky {R}osen paradox}. \textit{Physics Physique Fizika}. {\it 1}, 195--200.
\newblock \doi{10.1103/PhysicsPhysiqueFizika.1.195}.

\bibitem[{{Brans}(1988)}]{Brans1988Model}
{Brans, C.~H.} (1988) {B}ell's theorem does not eliminate fully causal hidden variables. \textit{International Journal of Theoretical Physics}. {\it 27}(2), 219--226.
\newblock \doi{10.1007/BF00670750}.

\bibitem[{{Calosi and Morganti}(2016)}]{Calosi2016}
{Calosi, C. and Morganti, M.} (2016) Humean supervenience, composition as identity and quantum wholes. \textit{Erkenntnis}. {\it 81}(6), 1173--1194.
\newblock \doi{10.1007/s10670-015-9789-z}.

\bibitem[{{Cao}(2003)}]{Cao2003}
{Cao, T.~Y.} (2003) Appendix: Ontological relativity and fundamentality -- is qft the fundamental theory? \textit{Synthese}. {\it 136}(1), 25--30.
\newblock \doi{10.1023/a:1024199931657}.

\bibitem[{{Chen}(2021)}]{Chen2021}
{Chen, E.~K.} (2021) Bell’s theorem, quantum probabilities, and superdeterminism In {\em The Routledge Companion to Philosophy of Physics},  Knox, E. and Wilson, A. (eds). Routledge. pp. 184--199.
\newblock \doi{10.4324/9781315623818}.

\bibitem[{{Ciepielewski et~al.}(2023){Ciepielewski, Okon, and Sudarsky}}]{Ciepielewski2023}
{Ciepielewski, G.~S., Okon, E. and Sudarsky, D.} (2023) On superdeterministic rejections of settings independence. \textit{The British Journal for the Philosophy of Science}. {\it 74}(2), 435--467.
\newblock \doi{10.1086/714819}.

\bibitem[{{Clauser et~al.}(1969){Clauser, Horne, Shimony, and Holt}}]{ClauserShimonyHolt1969}
{Clauser, J.~F., Horne, M.~A., Shimony, A. and Holt, R.~A.} (1969) Proposed experiment to test local hidden-variable theories. \textit{Physical Review Letters}. {\it 23}(15), 880--884.
\newblock \doi{10.1103/PhysRevLett.23.880}.

\bibitem[{{Crowther}(2019)}]{Crowther2019}
{Crowther, K.} (2019) When do we stop digging? {C}onditions on a fundamental theory of physics In {\em What Is Fundamental?},  Aguirre, A., Foster, B., and Merali, Z. (eds). The Frontiers Collection. Springer. pp. 123--133.
\newblock \doi{10.1007/978-3-030-11301-8\_13}.

\bibitem[{{Daley et~al.}(2022){Daley, Resch, and Spekkens}}]{Daley2022Adjudicating}
{Daley, P.~J., Resch, K.~J. and Spekkens, R.~W.} (2022) Experimentally adjudicating between different causal accounts of {B}ell-inequality violations via statistical model selection. \textit{Phys. Rev. A}. {\it 105}, 042220.
\newblock \doi{10.1103/PhysRevA.105.042220}.

\bibitem[{{Daley et~al.}(2025){Daley, Resch, and Spekkens}}]{Daley2024HanceReply}
{Daley, P.~J., Resch, K.~J. and Spekkens, R.~W.} (2025) Reply to ``{C}omment on `{E}xperimentally adjudicating between different causal accounts of {B}ell-inequality violations via statistical model selection' ''. \textit{Phys. Rev. A}. {\it 111}, 016201.
\newblock \doi{10.1103/PhysRevA.111.016201}.

\bibitem[{{Dawid et~al.}(2015){Dawid, Hartmann, and Sprenger}}]{Dawid2015}
{Dawid, R., Hartmann, S. and Sprenger, J.} (2015) The no alternatives argument. \textit{British Journal for the Philosophy of Science}. {\it 66}(1), 213--234.
\newblock \doi{10.1093/bjps/axt045}.

\bibitem[{{Dehmelt}(1989)}]{Dehmelt1989}
{Dehmelt, H.} (1989) Triton,... {E}lectron,... {C}osmon,...: {A}n infinite regression? \textit{Proceedings of the National Academy of Sciences}. {\it 86}(22), 8618--8619.
\newblock \doi{10.1073/pnas.86.22.8618}.

\bibitem[{{Donadi and Hossenfelder}(2022)}]{DonadiHossenfelder2022}
{Donadi, S. and Hossenfelder, S.} (2022) Toy model for local and deterministic wave-function collapse. \textit{Phys. Rev. A}. {\it 106}, 022212.
\newblock \doi{10.1103/PhysRevA.106.022212}.

\bibitem[{{Douglas}(2013)}]{douglas2013}
{Douglas, H.} (2013) The value of cognitive values. \textit{Philosophy of Science}. {\it 80}(5), 796--806.
\newblock \doi{10.1086/673716}.

\bibitem[{{Duerr and Fischer}(2025)}]{duerrfischer}
{Duerr, P.~M. and Fischer, E.} (2025) Rationally warranted promise: the virtue-economic account of pursuit-worthiness. \textit{Synthese}. {\it 206}.
\newblock \doi{10.1007/s11229-025-05077-5}.

\bibitem[{{Einstein et~al.}(1935){Einstein, Podolsky, and Rosen}}]{EinsteinPodolskyRosen1935}
{Einstein, A., Podolsky, B. and Rosen, N.} (1935) Can quantum-mechanical description of physical reality be considered complete? \textit{Physical Review}. {\it 47}(10), 777--780.
\newblock \doi{10.1103/PhysRev.47.777}.

\bibitem[{{Fischer}(2026)}]{Fischer2026}
{Fischer, E.} (2026) The pursuitworthiness of experiments. \textit{European Journal for Philosophy of Science}. {\it 16}(1), 5.
\newblock \doi{10.1007/s13194-025-00711-y}.

\bibitem[{{Fisher et~al.}(2021){Fisher, Gelfert, and Steinle}}]{Fisher2021}
{Fisher, G., Gelfert, A. and Steinle, F.} (2021) Exploratory models and exploratory modeling in science: {I}ntroduction. \textit{Perspectives on Science}. {\it 29}(4), 355--358.
\newblock \doi{10.1162/posc\_{e}\_00374}.

\bibitem[{{French and McKenzie}(2012)}]{French2012}
{French, S. and McKenzie, K.} (2012) Thinking outside the toolbox: {T}owards a more productive engagement between metaphysics and philosophy of physics. \textit{European Journal of Analytic Philosophy}. {\it 8}(1), 42--59.

\bibitem[{{Garza and Hance}(2025)}]{GarzaHance2025}
{Garza, A.~J. and Hance, J.~R.} (2025) Quantum-like correlations from local hidden-variable theories under conservation law.

\bibitem[{{Hance}(2024)}]{Hance2024}
{Hance, J.~R.} (2024) Counterfactual restrictions and {B}ell's theorem. \textit{Journal of Physics Communications}. {\it 8}(12), 122001.
\newblock \doi{10.1088/2399-6528/ad9b6d}.

\bibitem[{{Hance and Hossenfelder}(2022)}]{HanceHossenfelder2022}
{Hance, J.~R. and Hossenfelder, S.} (2022) What does it take to solve the measurement problem? \textit{Journal of Physics Communications}. {\it 6}(10), 102001.
\newblock \doi{10.1088/2399-6528/ac96cf}.

\bibitem[{{Hance and Hossenfelder}(2024)}]{Hance2024DaleyComment}
{Hance, J.~R. and Hossenfelder, S.} (2024) Comment on ``experimentally adjudicating between different causal accounts of bell-inequality violations via statistical model selection''. \textit{Phys. Rev. A}. {\it 109}, 026201.
\newblock \doi{10.1103/PhysRevA.109.026201}.

\bibitem[{{Hance et~al.}(2022){Hance, Hossenfelder, and Palmer}}]{HanceHossenfelderPalmer2022}
{Hance, J.~R., Hossenfelder, S. and Palmer, T.~N.} (2022) Supermeasured: Violating bell-statistical independence without violating physical statistical independence. \textit{Foundations of Physics}. {\it 52}, 81.
\newblock \doi{10.1007/s10701-022-00602-9}.

\bibitem[{{Hance et~al.}(2025){Hance, Palmer, and Rarity}}]{Hance_2025}
{Hance, J.~R., Palmer, T.~N. and Rarity, J.} (2025) Experimental tests of invariant set theory. \textit{Physica Scripta}. {\it 100}(6), 065123.
\newblock \doi{10.1088/1402-4896/add9d7}.

\bibitem[{{Hossenfelder}(2020)}]{Hossenfelder2020Superdeterminism}
{Hossenfelder, S.} (2020) Superdeterminism: A guide for the perplexed. \textit{arXiv preprint arXiv:2010.01324}.

\bibitem[{{Hossenfelder and Palmer}(2020)}]{HossenfelderPalmer2020}
{Hossenfelder, S. and Palmer, T.} (2020) Rethinking superdeterminism. \textit{Frontiers in Physics}. {\it 8}, 139.
\newblock \doi{10.3389/fphy.2020.00139}.

\bibitem[{{Huggett and W{\"u}thrich}(2018)}]{Huggett2018}
{Huggett, N. and W{\"u}thrich, C.} (2018) The (a)temporal emergence of spacetime. \textit{Philosophy of Science}. {\it 85}(December), 1190--1203.
\newblock \doi{10.1086/699723}.

\bibitem[{{Ismael and Schaffer}(2020)}]{IsmaelSchaffer2020}
{Ismael, J. and Schaffer, J.} (2020) Quantum holism: nonseparability as common ground. \textit{Synthese}. {\it 197}(10), 4131--4160.
\newblock \doi{10.1007/s11229-016-1201-2}.

\bibitem[{{Jammer}(1997)}]{Jammer1997}
{Jammer, M.} (1997) {\em Concepts of mass in classical and modern physics}. Courier Corporation.

\bibitem[{{Koslow}(1968)}]{Koslow1968}
{Koslow, A.} (1968) Mach's concept of mass: Program and definition. \textit{Synthese}. {\it 18}(2), 216--233.
\newblock \doi{10.1007/BF00413776}.

\bibitem[{{Kuhn}(1979)}]{Kuhn2024}
{Kuhn, T.~S.} (1979) {\em The essential tension: Selected studies in scientific tradition and change}. University of Chicago press.

\bibitem[{{Ladyman and Ross}(2007)}]{Ladyman2007}
{Ladyman, J. and Ross, D.} (2007) {\em Every thing must go: Metaphysics naturalized}. Oxford University Press.
\newblock \doi{10.1093/acprof:oso/9780199276196.001.0001}.

\bibitem[{{Le~Bihan}(2018)}]{LeBihan2018}
{Le~Bihan, B.} (2018) Priority monism beyond spacetime. \textit{Metaphysica}. {\it 19}(1), 95--111.
\newblock \doi{10.1515/mp-2018-0005}.

\bibitem[{{Lichtenstein}(2021)}]{Lichtenstein2021}
{Lichtenstein, E.~I.} (2021) (mis)understanding scientific disagreement: Success versus pursuit-worthiness in theory choice. \textit{Studies in History and Philosophy of Science Part A}. {\it 85}(C), 166--175.
\newblock \doi{10.1016/j.shpsa.2020.10.005}.

\bibitem[{{Maudlin}(2007)}]{Maudlin2007}
{Maudlin, T.} (2007) {\em The metaphysics within physics}. Oxford University Press.
\newblock \doi{10.1093/acprof:oso/9780199218219.001.0001}.

\bibitem[{{Maudlin}(2019)}]{Maudlin2019}
{Maudlin, T.} (2019) Comment on ``electrons don't think''. Blog post on \textit{BackRe(Action)}. Retrieved December 1, 2019.

\bibitem[{{Maxwell}(1920)}]{Maxwell1991}
{Maxwell, J.~C.} (1920) {\em Matter and motion}. London MacMillan.

\bibitem[{{McMullin}(1982)}]{Mcmullin1982}
{McMullin, E.} (1982) Values in science. PSA: Proceedings of the biennial meeting of the philosophy of science association. Cambridge University Press. pp. 2--28.
\newblock \doi{10.1086/psaprocbienmeetp.1982.2.192409}.

\bibitem[{{Morganti}(2009)}]{Morganti2009RelationalHolism}
{Morganti, M.} (2009) A new look at relational holism in quantum mechanics. \textit{Journal for General Philosophy of Science}. {\it 40}(2), 297--314.
\newblock \doi{10.1007/s10838-009-9095-5}.

\bibitem[{{Morganti}(2020{\natexlab{a}})}]{Morganti2020a}
{Morganti, M.} (2020{\natexlab{a}}) Fundamentality in metaphysics and the philosophy of physics. part i: Metaphysics. \textit{Philosophy Compass}. {\it 15}(7), e12690.
\newblock \doi{10.1111/phc3.12690}.

\bibitem[{{Morganti}(2020{\natexlab{b}})}]{Morganti2020b}
{Morganti, M.} (2020{\natexlab{b}}) Fundamentality in metaphysics and the philosophy of physics. part ii: The philosophy of physics. \textit{Philosophy Compass}. {\it 15}(10), e12703.
\newblock \doi{10.1111/phc3.12703}.

\bibitem[{{Morganti and Calosi}(2021)}]{Morganti2021}
{Morganti, M. and Calosi, C.} (2021) Interpreting quantum entanglement: {S}teps towards coherentist quantum mechanics. \textit{British Journal for the Philosophy of Science}. {\it 72}(3), 865--891.
\newblock \doi{10.1093/bjps/axy064}.

\bibitem[{{Palmer}(2009)}]{Palmer2009}
{Palmer, T.} (2009) The invariant set postulate: a new geometric framework for the foundations of quantum theory and the role played by gravity. \textit{Proceedings of the Royal Society A}. {\it 465}, 3187--3207.
\newblock \doi{10.1098/rspa.2009.0080}.

\bibitem[{{Palmer}(2014)}]{Palmer2014}
{Palmer, T.} (2014) Lorenz, {G}odel and {P}enrose: new perspectives on determinism and causality in fundamental physics. \textit{Contemporary Physics}. {\it 55}(3), 1--29.
\newblock \doi{10.1080/00107514.2014.908624}.

\bibitem[{{Palmer}(2016)}]{Palmer2016}
{Palmer, T.} (2016) Invariant set theory.

\bibitem[{{Palmer}(2019)}]{Palmer2019}
{Palmer, T.} (2019) Bell inequality violation with free choice and local causality on the invariant set.

\bibitem[{{Palmer}(2024)}]{Palmer2024}
{Palmer, T.} (2024) Superdeterminism without conspiracy. \textit{Universe}. {\it 10}(1).
\newblock \doi{10.3390/universe10010047}.

\bibitem[{{Palmer}(2026)}]{palmer2025testingquantummechanicsquantum}
{Palmer, T.} (2026) Rational quantum mechanics: Testing quantum theory with quantum computers. \textit{Proceedings of the National Academy of Sciences}. {\it 123}(12), e2523350123.
\newblock \doi{10.1073/pnas.2523350123}.

\bibitem[{{Paul}(2002)}]{Paul2002}
{Paul, L.~A.} (2002) Logical parts. \textit{No\^{u}s}. {\it 36}(4), 578--596.
\newblock \doi{10.1111/1468-0068.00402}.

\bibitem[{{Pearl}(2009)}]{pearl2009causality}
{Pearl, J.} (2009) {\em Causality}. Cambridge University Press. Cambridge. second edition.

\bibitem[{{Rovelli and Vidotto}(2014)}]{Rovelli2014}
{Rovelli, C. and Vidotto, F.} (2014) {\em Covariant Loop Quantum Gravity}. Cambridge University Press.
\newblock \doi{10.1017/CBO9781107706910}.

\bibitem[{{Schaffer}(2009)}]{Schaffer2009}
{Schaffer, J.} (2009) Spacetime the one substance. \textit{Philosophical Studies}. {\it 145}, 131--148.
\newblock \doi{10.1007/s11098-009-9386-6}.

\bibitem[{{Schaffer}(2010)}]{Schaffer2010Monism}
{Schaffer, J.} (2010) Monism: The priority of the whole. \textit{The Philosophical Review}. {\it 119}(1), 31--76.
\newblock \doi{10.1215/00318108-2009-025}.

\bibitem[{{Schaffer}(2024)}]{Schaffer2024Monism}
{Schaffer, J.} (2024) Monism In {\em The Stanford Encyclopedia of Philosophy},  Zalta, E.~N. and Nodelman, U. (eds). Metaphysics Research Lab, Stanford University. {F}all 2024 edition.

\bibitem[{{Sen and Valentini}(2020{\natexlab{a}})}]{SenValentini2020}
{Sen, I. and Valentini, A.} (2020{\natexlab{a}}) Superdeterministic hidden-variables models i: Non-equilibrium and signalling. \textit{Proceedings of the Royal Society A}. {\it 476}, 20200212.
\newblock \doi{10.1098/rspa.2020.0212}.

\bibitem[{{Sen and Valentini}(2020{\natexlab{b}})}]{SenValentini2020b}
{Sen, I. and Valentini, A.} (2020{\natexlab{b}}) Superdeterministic hidden-variables models ii: Conspiracy. \textit{Proceedings of the Royal Society A}. {\it 476}, 20200214.
\newblock \doi{10.1098/rspa.2020.0214}.

\bibitem[{{Shaw}(2022)}]{Shaw2022}
{Shaw, J.} (2022) On the very idea of pursuitworthiness. \textit{Studies in History and Philosophy of Science Part A}. {\it 91}(C), 103--112.
\newblock \doi{10.1016/j.shpsa.2021.11.016}.

\bibitem[{{t~Hooft}(2016)}]{tHooft2016cellular}
{t~Hooft, G.} (2016) {\em The cellular automaton interpretation of quantum mechanics}. Fundamental Theories of Physics. Springer Nature.
\newblock \doi{10.1007/978-3-319-41285-6}.

\bibitem[{{Tahko}(2023)}]{Tahko2023Fundamentality}
{Tahko, T.~E.} (2023) {Fundamentality} In {\em The {Stanford} Encyclopedia of Philosophy},  Zalta, E.~N. and Nodelman, U. (eds). Metaphysics Research Lab, Stanford University. {W}inter 2023 edition.

\bibitem[{{Teller}(1986)}]{Teller1986RelationalHolism}
{Teller, P.} (1986) Relational holism and quantum mechanics. \textit{British Journal for the Philosophy of Science}. {\it 37}(1), 71--81.
\newblock \doi{10.1093/bjps/37.1.71}.

\bibitem[{{Teller}(1989)}]{Teller1989RelativityHolismBell}
{Teller, P.} (1989) Relativity, relational holism, and the bell inequalities In {\em Philosophical Consequences of Quantum Theory: Reflections on Bell's Theorem},  Cushing, J.~T. and McMullin, E. (eds). University of Notre Dame Press. Notre Dame, IN. pp. 208--223.

\bibitem[{{Thomson and Tait}(2022)}]{Thomson2022}
{Thomson, W. and Tait, P.~G.} (2022) {\em Treatise on natural philosophy}. Cambridge, University Press.
\newblock \doi{10.1017/CBO9780511703928}.

\bibitem[{{Waegell and McQueen}(2025)}]{Waegell2025}
{Waegell, M. and McQueen, K.~J.} (2025) From statistical dependence to the space of possible superdeterministic theories. \textit{European Journal for Philosophy of Science}. {\it 15}(4), 56.
\newblock \doi{10.1007/s13194-025-00693-x}.

\bibitem[{{Wharton and Argaman}(2020)}]{Wharton2019Reformulations}
{Wharton, K.~B. and Argaman, N.} (2020) Colloquium: {B}ell's theorem and locally mediated reformulations of quantum mechanics. \textit{Rev. Mod. Phys.} {\it 92}, 021002.
\newblock \doi{10.1103/RevModPhys.92.021002}.

\bibitem[{{Zurek}(1991)}]{Zurek1991Decoherence}
{Zurek, W.~H.} (1991) Decoherence and the transition from quantum to classical. \textit{Physics Today}. {\it 44}(10), 36--44.
\newblock \doi{10.1063/1.881293}.

\end{thebibliography}
\end{document}